\documentclass[aps,twocolumn,a4paper,floatfix,superscriptaddress,pra,longbibliography]{revtex4-2}
\usepackage[toc,page]{appendix}
\usepackage{braket}
\usepackage{epsfig,graphicx,times}
\usepackage{amstext}
\usepackage{amsmath}
\usepackage{amssymb}
\usepackage{mathrsfs}
\usepackage{dcolumn}
\usepackage{bm}
\usepackage[tight]{subfigure}
\usepackage[colorlinks,linkcolor=blue,anchorcolor=blue,citecolor=blue,urlcolor=blue]{hyperref}
\usepackage{color}
\usepackage{siunitx}
\usepackage{appendix}
\usepackage{placeins}
\usepackage{textcomp}
\usepackage{bbold}
\usepackage{float}
\usepackage{verbatim}
\usepackage{minitoc}
\usepackage[dvipsnames]{xcolor}

\usepackage{soul}
\usepackage[framemethod=tikz]{mdframed}

\begin{document}
	
\title{Feedback Control and Neural-Network-Based Prediction of Quantum Entanglement in Fano-mirror cavity optomechanics}
	
\author{Ya-Feng Jiao}
\email{yfjiao@zzuli.edu.cn}
\affiliation{School of Electronics and Information, Henan Key Laboratory of Information Functional Materials and Sensing Technology, Zhengzhou University of Light Industry, Zhengzhou 450001, P.R.China}
\affiliation{Academy for Quantum Science and Technology, Zhengzhou University of Light Industry, Zhengzhou 450001, P.R.China}

\author{Ze-Yu Zhu}
\affiliation{School of Electronics and Information, Henan Key Laboratory of Information Functional Materials and Sensing Technology, Zhengzhou University of Light Industry, Zhengzhou 450001, P.R.China}

\author{Jie Wang}
\affiliation{College of Science, NUDT, Changsha 410073, China}

\author{Qing-Shou Tan}
\affiliation{Key Laboratory of Hunan Province on Information Photonics and Free space Optical Communication, College of Physics and Electronics, Hunan Institute of Science and Technology, Yueyang 414000, China}

\author{Wen-Qiang Fang}
\affiliation{School of Electronics and Information, Henan Key Laboratory of Information Functional Materials and Sensing Technology, Zhengzhou University of Light Industry, Zhengzhou 450001, P.R.China}

\author{Le-Man Kuang}
\email{lmkuang@hunnu.edu.cn}
\affiliation{Key Laboratory of Low-Dimensional Quantum Structures and Quantum Control of Ministry of Education, \\
Department of Physics and Synergetic Innovation Center for Quantum Effects and Applications, \\ Hunan Normal University, Changsha 410081, China}
\affiliation{Academy for Quantum Science and Technology, Zhengzhou University of Light Industry, Zhengzhou 450001, P.R.China}

\author{Hui Jing}
\email{jinghui@hunnu.edu.cn}
\affiliation{Key Laboratory of Low-Dimensional Quantum Structures and Quantum Control of Ministry of Education, \\
Department of Physics and Synergetic Innovation Center for Quantum Effects and Applications, \\ Hunan Normal University, Changsha 410081, China}
\affiliation{Academy for Quantum Science and Technology, Zhengzhou University of Light Industry, Zhengzhou 450001, P.R.China}
	
\date{\today}
	
\begin{abstract}
Quantum entanglement involving the centre-of-mass motion of macroscopic-scale objects plays an essential role in both fundamental science and emerging quantum technologies, yet generating and verifying such macroscopic nonclassical states remain formidable challenges. Here, we report a theoretical scheme to effectively generate, manipulate, and predict bipartite optomechanical entanglement based on a Fano-mirror cavity optomechanical system with a coherent feedback loop, where the cavity and Fano optical modes couple through coherent and dissipative interactions. We show that the Fano mirror forms a closed tripartite interaction loop with the cavity and mechanical modes, whereby the quantum interference between direct and indirect interaction pathways substantially enhances the cavity-mechanical entanglement and gives rise to a robust Fano-mechanical entanglement. More importantly, we find that coherent feedback combined with an appropriate optical frequency mismatch can further enhance the entanglement, and a trained supervised feedforward neural network can be used to predict the entanglement magnitude from the readily measurable optical signals, relaxing the requirement of full quantum-state tomography. Our work, establishing an all-optical and experimentally feasible approach to generate, manipulate, and characterize the macroscopic optomechanical entanglement, holds great potential for applications in quantum information processing.
\end{abstract}	
	
\maketitle

\section{Introduction}
Quantum entanglement, as a striking feature of quantum mechanics that allows for nonclassical correlations between distant quantum systems~\cite{Horodecki2009RMP}, is an appealing resource that underpins a wide range of emerging quantum technologies spanning from quantum communication~\cite{Zhang2019NSR}, quantum sensing~\cite{Degen2017RMP} to quantum networking~\cite{Kimble2008Nature}. Particularly, the creation and verification of entangled states involving macroscopic-scale objects have attracted intense interests over the past decades, since such macroscopic nonclassical states not only offer a promising testbed for exploring the fundamentals of quantum theory like the Bell test~\cite{Hofer2016PRL,Simon2018PRL} and the classical-quantum boundary~\cite{Leggett2002JPCM}, but also constitute indispensable resources for advancing many disruptive quantum technologies beyond the classical limits~\cite{Dowling2003PTRSA}. So far, macroscopic entanglement has been demonstrated in various types of quantum system~\cite{Raimond2001RMP,Pan2012RMP,Dai2016NP,Luo2022PRL,Riedinger2018Nature}, including solid-state spin ensembles~\cite{Simmons2011Nature,Klimov2015SA}, NV-centers~\cite{Neumann2008Science}, superconducting qubits~\cite{Campagne2018PRL,Kurpiers2018Nature}, Bose-Einstein Condensate (BEC)~\cite{Lettner2011PRL}, and microwave~\cite{Korppi2018Nature} or electromechanical~\cite{Barzanjeh2019Nature} devices. Recently, cavity optomechanical (COM) system~\cite{Aspelmeyer2014RMP}, leveraging the radiation-pressure-induced interplay between confined optical fields and mechanical oscillators, provides a versatile platform for investigating a broad range of macroscopic quantum phenomena~\cite{Xiong2015SCPMA,Barzanjeh2022NP}, such as quantum synchronization of mechanical oscillation~\cite{Santos2017PRL,Li2022PRL,Lai2025NC}, mechanical dynamical Casimir effect~\cite{Jiang2025PRA}, photon~\cite{Rabl2011PRL,Liao2013PRA} or phonon~\cite{Xie2017PRA,Zheng2019PRA} blockade, quantum phase transitions~\cite{Lu2018PRAPP,Fruchart2021Nature,Zhu2024PRL}, quantum chaos~\cite{Zhu2019PRA,ZhangDW2021PRA}, optomechanical squeezing~\cite{Aggarwal2020NP,Huang2023PRA,Jiao2026AQT}, and phonon lasing~\cite{Kuang2023NP}, to name a few. Notably, by exploiting the nonlinear light-motion interaction in COM system, remarkable progress has been made toward observing quantum entanglement between optical propagating fields~\cite{Barzanjeh2019Nature,Chen2020NC}, between light and mechanical vibration~\cite{Vitali2007PRL,Palomaki2013Science,Deplano2026Science}, as well as between spatially separated mechanical oscillators~\cite{Kotler2021Science,Mercier2021Science}. Nevertheless, such macroscopic entanglement is typically vulnerable to environmental disturbances, and the observation of strong macroscopic entanglement in COM device is hampered by the weak photon-phonon interaction and the system's limited stability. To prepare highly entangled states in COM systems, extensive efforts have been made to develop strategies for COM entanglement enhancement and protection, including dark-mode~\cite{Huang2022PRA2,Lai2022PRR} or time-reversal-symmetry~\cite{Jiao2020PRL} breaking, reservoir engineering~\cite{Wang2013PRL}, coherent feedback control~\cite{Li2017PRA}, and synthetic magnetism~\cite{Lai2022PRL}. On the other hand, the experimental characterization of COM entanglement also remains a formidable task, because certifying COM entanglement generally requires the reconstruction of the full covariance matrix of the involved modes~\cite{Vitali2007PRL,Palomaki2013Science,Deplano2026Science,Kotler2021Science,Mercier2021Science}, which in turn demands simultaneous access to all the optical and mechanical quadratures. In experiments, the mechanical quadratures are difficult to measure directly and the reconstruction is sensitive to detection noise and parameter fluctuations. Machine learning, as a powerful tool for tackling complex problems, provides an effective means to circumvent this difficulty~\cite{Carleo2019RMP}. For example, a recent theoretical scheme demonstrates that a supervised feedforward neural network, whose approximation capability is guaranteed by the universal approximation theorem~\cite{Hornik1991NN}, can be used to predict strong phonon blockade in a quadratically coupled COM system from the directly detected optical signals~\cite{Zeng2021APL}, without the need to access the mechanical mode. This result indicates a promising route for estimating nonclassical properties of mechanical systems from readily measurable optical data.

\begin{figure*}[htbp]
	\centering
	\includegraphics[width=0.98\textwidth]{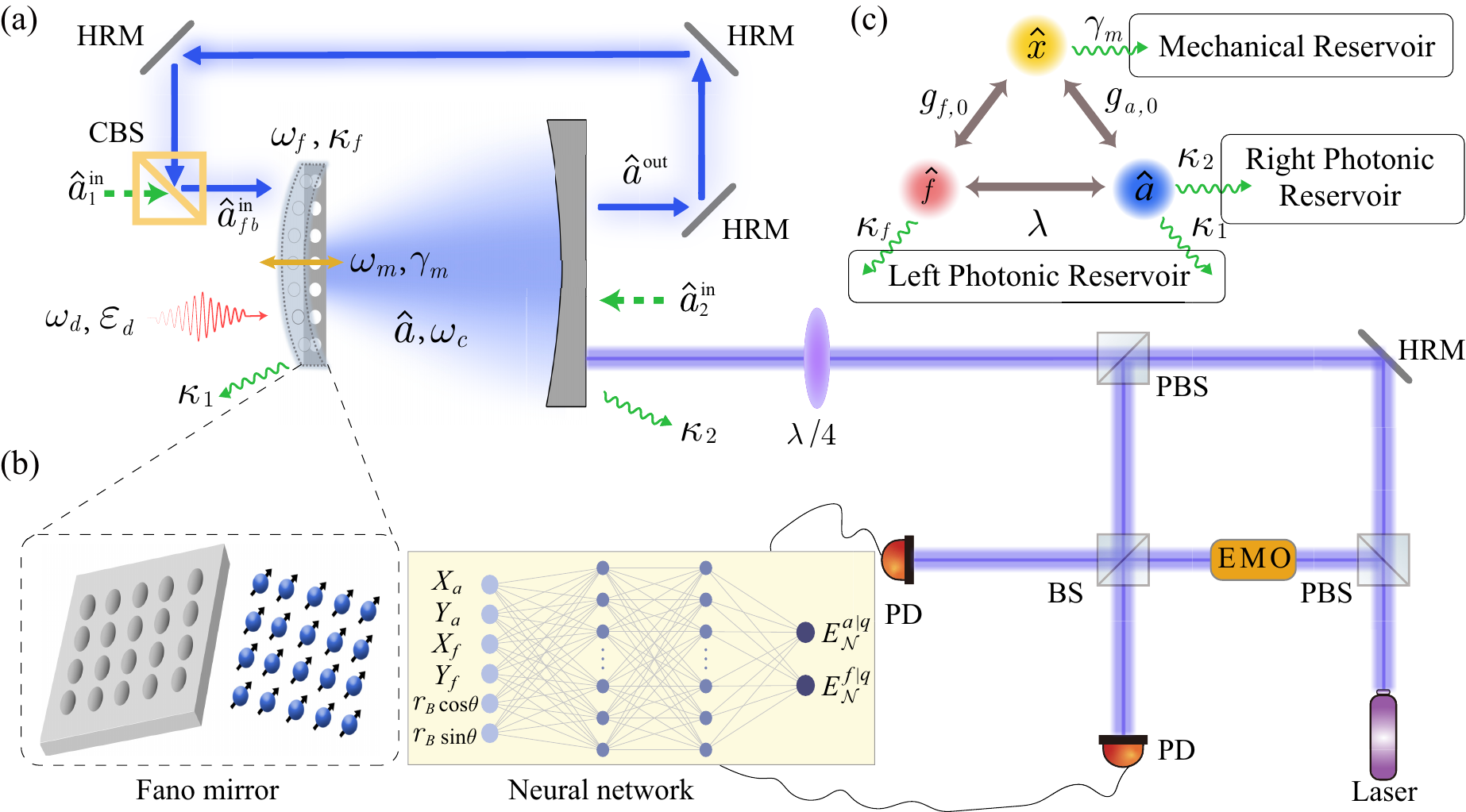}
\caption{(a) Schematics of a Fano-mirror COM system coupled with a coherent feedback loop and a homodyne detection setup. The input of the feedforward neural network is the optical output-field quadratures measured by the homodyne detection, and the corresponding output is the logarithmic negativity $E_{\mathcal{N}}^{a|q}$ and $E_{\mathcal{N}}^{f|q}$. (b) Possible realizations of the Fano mirror: a dielectric membrane patterned with the designed subwavelength PhC structures, or a membrane doped with an ensemble of atoms arranged in a two-dimensional array, which collectively behaves as a single bosonic mode. (c) Closed-loop coupling diagram of the three mode in the fundamental nonlinear configuration described by Hamiltonian (\ref{eq:Hamiltonian}).}\label{Fig1}
\end{figure*}

In parallel, Fano mirrors have recently emerged as a flexible tool to efficiently improve single-photon COM coupling and sideband resolution~\cite{Miroshnichenko2010RMP,Limonov2017NP}, opening up new possibilities for exploring nonclassical phenomena in the quantum nonlinear regime. In practice, Fano mirrors can be fabricated by patterning subwavelength gratings~\cite{Chang2012AOP} or photonic-crystal (PhC) structures~\cite{Zhou2014PQE} onto dielectric membranes, employing semiconducting monolayers~\cite{Back2018PRL,Scuri2018PRL}, or arranging atoms in two-dimensional arrays within optical lattices~\cite{Bettles2016PRL,Shahmoon2017PRL,Wild2018PRL}. Very recently, Fano-mirror COM system has been experimentally formed by replacing one of the cavity end-mirror with a Fano mirror~\cite{Manjeshwar2023OE,Du2026QST,Monsel2024PRA,Du2025PRA}. In this architecture, the Fano mirror can simultaneously support a mechanical vibration mode and a guided optical resonance, i.e., the so-called Fano mode, whose interference with the cavity mode gives rise to a Fano-type optical response of the mirror. The cavity and Fano modes are coupled not only through the spatial overlap of their optical fields, but also through their common decay into the same photonic reservoir~\cite{Claudiu2019PRL}, while both optical modes can be dispersively coupled to the mechanical mode. These unique features simultaneously enable a strong single-photon COM coupling~\cite{Du2026QST} and a severe reduction of the effective optical linewidth~\cite{Monsel2024PRA,Du2025PRA}, thereby allowing the Fano-mirror COM sytem to engineer diverse nonlinear optomechanical effects in the resolved-sideband regime. However, up to now, the potential of Fano-type quantum interference for generating and enhancing COM entanglement has not yet been reported.

Inspired by these recent advances, we here investigate how to achieve feedback-assited enhancement and neural-network-based prediction of bipartite quantum entanglement in a Fano-mirror COM system. Specifically, we show that by replacing one cavity end-mirror of the COM system with a Fano mirror, the cavity, Fano, and mechanical modes form a closed tripartite interaction loop, wherein the quantum interference between the direct and indirect interaction pathways allows us both to markedly improve the original cavity-mechanical entanglement and to simultaneously generate a robust Fano-mechanical entanglement. More interestingly, we further show that when introducing a coherent feedback loop to the COM system, it renormalizes the effective cavity damping and detuning and reshapes the noise correlations between the cavity and Fano modes. The resulting modification of the input-noise correlations further strengthens the quantum interference in the closed loop, leading to an additional enhancement of both the cavity-mechanical and Fano-mechanical entanglement. Finally, we demonstrate that a trained feedforward neural network provides a reliable tool to predict the generated entanglement from the readily measurable optical signals obtained via balanced homodyne detection, which bypasses the full quantum-state tomography and is robust against slight parameter fluctuations. Overall, our work establishes an all-optical approach to generate, enhance, and characterize macroscopic COM entanglement, with great potentials for applications ranging from entanglement-enabled quantum sensing to quantum networking.

In Sec.\,\ref{secII}, we present the theoretical model and derive the system dynamics, together with the quantitative measures of COM entanglement. In Sec.\,\ref{secIII}, we present the numerical simulations of the entanglement enhancement under various controlling parameters, and analyze the underlying physical mechanism. In Sec.\,\ref{secIV}, we demonstrate the machine-learning-based prediction of the bipartite entanglement. Finally, a brief summary is given in Sec.\,\ref{secV}.

\section{Theoretical Model}\label{secII}
As illustrated in Fig.\,\ref{Fig1}(a), we consider a hybrid COM system formed by a Fabry-P\'{e}rot-type (FP) cavity incorporating a coherent feedback loop and a homodyne detection setup. In the FP cavity, as shown in Fig.\,\ref{Fig1}(b), the left cavity boundary is defined by a Fano mirror with optical decay rate $\kappa_{1}$, which can be experimentally realized by employing, e.g., a suspended dielectric membrane patterned with the designed subwavelength PhC structures~\cite{Zhou2014PQE,Manjeshwar2023OE} or two-dimensional atomic arrays~\cite{Bettles2016PRL,Shahmoon2017PRL,Wild2018PRL}. This Fano mirror is considered to be movable and supports a mechanical vibration mode described by dimensionless displacement $\hat{x}$ and momentum $\hat{p}$ with fundamental frequency $\Omega_m$ and mechanical damping rate $\gamma_{m}$, as well as a guided optical mode $\hat{f}$ with resonance frequency $\omega_f$ and optical decay rate $\kappa_{f}$, hereafter referred to as the Fano mode. The right cavity boundary is defined by a normal mirror (e.g., a distributed Bragg reflector) with optical decay rate $\kappa_{2}$. The intracavity optical field, characterized by a cavity mode $\hat{a}$ with resonance frequency $\omega_a$, is confined between these two mirrors. Note that since the cavity mode decays through both the left Fano mirror and the right normal mirror, its total decay rate is given by $\kappa_a = \kappa_1 + \kappa_2$. The mechanical vibration mode could couple simultaneously to both the cavity and Fano optical modes via dispersive optomechanical interactions with single-photon coupling strengths of $g_{a,0}$ and $g_{f,0}$, respectively. Here, the cavity-vibration optomechanical interaction is induced by the optical-radiation-pressure mediated coupling between the cavity field and the out-of-plane displacement of the Fano mirror~\cite{Manjeshwar2023OE}. In contrast, the Fano-vibration optomechanical interaction originates from the strain-induced deformations of the PhC structures, where the in-plane mechanical strain is generated by the out-of-plane displacement of the Fano mirror~\cite{Manjeshwar2023OE}. Meanwhile, the cavity and Fano optical modes are strongly coupled via both dispersive and dissipative interactions. The dispersive interaction originates from the spatial overlap of their optical fields and is characterized by a coherent coupling strength $J$~\cite{Cernotik2019QST}. The dissipative interaction arises from the collective dissipation of the cavity and Fano modes into the same electromagnetic environment through the left Fano mirror, giving rise to an effective dissipative coupling of strength $\kappa_{1f}=\sqrt{\kappa_1\kappa_f}$~\cite{Claudiu2019PRL}. As a result, the overall optical coupling between the cavity and Fano modes is described by a complex coupling strength $\lambda=J-i\kappa_{1f}$, and the optical and mechanical modes thus form a closed tripartite interaction loop [see Fig.\,\ref{Fig1}(c)]. In addition, we further assume that the cavity is driven through the left Fano mirror by a coherent driving field with amplitude $\varepsilon_{d}$ and frequency $\omega_d$, where $\varepsilon_{d}$ is related to the input laser power $P_{d}$ by $|\varepsilon_{d}|=\sqrt{2\kappa_{1}P_{d}/\hbar\omega_{d}}$. Accordingly, the Hamiltonian of such a Fano-mirror COM system, in the frame rotating at the driving frequency $\omega_{d}$, is given by ($\hbar=1$)
\begin{align}
\hat{H}=&\Delta_{a}\hat{a}^{\dagger}\hat{a}+\Delta_{f}\hat{f}^{\dagger}\hat{f}+\frac{\Omega_{m}}{2}(\hat{x}^{2}+\hat{p}^{2})+\lambda(\hat{a}^{\dagger}\hat{f}+\hat{f}^{\dagger}\hat{a}) \notag \\
&+(g_{a,0}\hat{a}^{\dagger}\hat{a}+g_{f,0}\hat{f}^{\dagger}\hat{f})\hat{x}+i\varepsilon_{d}(\hat{a}^{\dagger} - \hat{a}),
\label{eq:Hamiltonian}
\end{align}
where $\Delta_a=\omega_a-\omega_d$ and $\Delta_f=\omega_f-\omega_d$ are the optical detunings of the cavity and Fano modes from the driving field, respectively, and the optical and mechanical operators satisfy the canonical commutation relations $[\hat{a},\hat{a}^{\dagger}]=[\hat{f},\hat{f}^{\dagger}]=1$ and $[\hat{x},\hat{p}]=i$. In Eq.\,(\ref{eq:Hamiltonian}), the first three terms correspond to the free Hamiltonians of the cavity, Fano, and mechanical modes, respectively. The fourth term describes the optical coupling between the cavity and Fano modes, while the fifth term accounts for the Hamiltonian of the cavity- and Fano-optomechanical interactions. The last term is the Hamiltonian of the coherent driving field.
	
Based on the system Hamiltonian, the quantum Langevin equations (QLEs) governing the system dynamics can be derived by incorporating mode dissipations and input noises, i.e.,
\begin{align}
\dot{\hat{x}}=&\,\Omega_{m}\hat{p}, \notag\\
\dot{\hat{p}}=&-\Omega_{m}\hat{x}-\gamma_{m}\hat{p}-g_{a,0}\hat{a}^{\dagger}\hat{a}-g_{f,0}\hat{f}^{\dagger}\hat{f}+\hat{\xi},\notag\\
\dot{\hat{a}}=&\!-\!(i\Delta_a\!+\!\kappa_a)\hat{a}\!-\!ig_{a,0}\hat{a}\hat{x}\!-\!i\lambda\hat{f}\!+\!\varepsilon_{d}\!+\!\sum_{j=1,2}\!\sqrt{2\kappa_j}\hat{a}_j^{\mathrm{in}},\notag\\
\dot{\hat{f}}=&-(i\Delta_f+\kappa_f)\hat{f}-ig_{f,0}\hat{f}\hat{x}-i\lambda\hat{a}+\sqrt{2\kappa_f}\hat{a}_1^{\mathrm{in}},
\label{eq:QLEs}
\end{align}
where $\hat a_{1}^{\mathrm{in}}$ and $\hat a_{2}^{\mathrm{in}}$ denote the vacuum input noise operators associated with the left and right photonic reservoirs, respectively, which couple into the cavity and Fano modes through the left Fano and right normal mirrors. These vacuum input noise operators have zero mean and are characterized by the following nonvanishing correlation function~\cite{Zoller2000book}: $\langle \hat{a}_{j}^{\textrm{in}}(t)\hat{a}_{j}^{\textrm{in},\dagger}(t^{\prime})\rangle=\delta(t-t')$, with $j=1,2$. The noise operator $\hat{\xi}$ describes the stochastic Brownian force acting on the mechanical vibration mode, which arises from its coupling to a thermal reservoir at temperature $T$. Although the intrinsic correlation of $\hat{\xi}$ is generally non-Markovian, for a high-$Q$ mechanical oscillator, i.e., $Q_m = \Omega_m/\gamma_m \gg 1$, a Markovian approximation can be safely applied. This simplifies the correlation function to the delta-correlated form~\cite{Zoller2000book}: $\langle\hat{\xi}(t)\hat{\xi}(t')\rangle \simeq \gamma_m(2\bar{n}_m+1)\delta(t-t')$, where $\bar{n}_m = [\exp(\hbar\Omega_m/k_B T)-1]^{-1}$ denotes the mean thermal phonon occupancy, with $k_B$ the Boltzmann constant.

Because of the nonlinear dynamical terms introduced by radiation-pressure interactions in QLEs\,(\ref{eq:QLEs}), the system dynamics becomes difficult to be directly solved. To circumvent this difficulty, one can linearize the QLEs\,(\ref{eq:QLEs}) under strong optical driving condition by expanding each operator as a sum of its steady-state mean value and a quantum fluctuation around it, i.e., $\hat{a}=\alpha_{s}+\delta\hat{a}$, $\hat{f}=f_{s}+\delta\hat{f}$, $\hat{x}=x_{s}+\delta\hat{x}$, $\hat{p}=p_{s}+\delta\hat{p}$. Substituting this linearization ansatz into the QLEs\,(\ref{eq:QLEs}), one can obtain a set of nonlinear algebraic equations for the steady-state mean values and a set of linearized QLEs for the fluctuation operators. The steady-state mean values can then be determined by solving the algebraic equations, yielding
\begin{align}
p_{s}=&0,\notag\\
x_s=&-\frac{1}{\Omega_m}(g_{a,0}|a_s|^2+g_{f,0}|f_s|^2),\notag \\
a_s=&\frac{(i\bar\Delta_{f}+\kappa_f)\varepsilon_{d}}{(i\bar\Delta_{a}+\kappa_a)(i\bar\Delta_{f}+\kappa_f)+\lambda^2},\notag \\
f_s=&\frac{-i\lambda\varepsilon_{d}}{(i\bar\Delta_{a}+\kappa_a)(i\bar\Delta_{f}+\kappa_f)+\lambda^2},
\label{eq:SS-meanvalue}
\end{align}
where $\bar\Delta_{a}=\Delta_a+g_{a,0}x_s$ and $\bar\Delta_{f}=\Delta_f+g_{f,0}x_s$ are the effective optical detunings including
the radiation-pressure-induced frequency shift. Correspondingly, by substituting these steady-state mean values into the QLEs (\ref{eq:QLEs}) and neglecting the higher-order fluctuation terms, the linearized QLEs are obtained as
\begin{align}
\delta\dot{\hat{x}}=&\Omega_m\delta\hat{p},\notag\\
\delta\dot{\hat{p}}=&\!-\!\Omega_m\delta\hat{x}\!-\!\gamma_m\delta\hat{p}
-G_a(\delta\hat{a}\!+\!\delta\hat{a}^\dagger)-G_f(\delta\hat{f}\!+\!\delta\hat{f}^\dagger)
\!+\!\hat{\xi},\notag \\
\delta\dot{\hat{a}}=&-(i\bar\Delta_{a}+\kappa_a)\delta\hat{a}-iG_a\delta\hat{x}-i\lambda\delta\hat{f}
+\sum_{j=1,2}\sqrt{2\kappa_j}\hat{a}_j^{\mathrm{in}},\notag \\	
\delta\dot{\hat{f}}=&-(i\bar\Delta_{f}+\kappa_f)\delta\hat{f}-iG_f\delta\hat{x}-i\lambda\delta\hat{a}
+\sqrt{2\kappa_f}\hat{a}_1^{\mathrm{in}},
\label{eq:linQLEs}
\end{align}
where $G_a = g_{a,0}\bar{a}$ and $G_f = g_{f,0}\bar{f}$ are the effective COM coupling strengths associated with the cavity and Fano modes, respectively. Because the magnitudes of both $G_a$ and $G_f$ are proportional to the steady-state field amplitudes, the corresponding effective COM interactions could be significantly enhanced under strong optical driving. Moreover, in our following analysis, $G_a$ and $G_f$ are taken to be real without loss of generality, which can be achieved by choosing an appropriate phase reference for the steady-state mean values.

Furthermore, when introducing a double-sided coherent feedback loop to the Fano-mirror COM system, the optical output field transmitted from the left-hand Fano mirror can be sent back into the cavity through the right-hand normal mirror via a unidirectional traveling-wave field. In practice, the feedback loop is constructed with several highly reflective mirrors (HRMs) and a controllable beam splitter (CBS) with tunable reflection coefficient $r_{B}$, as shown in Fig.\,\ref{Fig1}(b). Due to field mixing in the CBS, the new input field $\hat{a}_{2,fb}^{\rm in}$ entering the cavity from the right-hand normal mirror is modified into a superposition of the original input field $\hat{a}_{2}^{\rm in}$ and the feedback-returned output field $\hat{a}_{1}^{\rm out}$. By using the standard input-output formula, we have
\begin{align}
\hat{a}_{1}^{\rm out}=&\sqrt{2\kappa_1} \delta\hat{a} - \hat{a}_{1}^{\rm in}, \notag \\
\hat{a}_{2,fb}^{\rm in}=&r_{B}e^{i \theta}\,\hat{a}_{1}^{\rm out} +t_{B}\hat{a}_{2}^{\rm in,\prime},
\label{eq:io}
\end{align}
where $r_B$ and $t_B$ are the reflection and transmission coefficients of the CBS, with $r_B^2+t_B^2=1$. The additional phase shift $\theta$ describes the phase delay accumulated by the reflection of the output field in the feedback loop, which is defined as $\theta =2\pi nl/\lambda_{d}$, with $n$ ($l$) the refractive index (length) of the loop and $\lambda_{d}$ the light wavelength. It should be noted that, in deriving Eq.\,(\ref{eq:io}), the reflection coefficient $r_B$ and the additional phase shift $\theta$ are defined in an effective way, such that all possible optical losses (e.g., those induced by the finite transmission of HRMs and the dissipation of CBS) and propagation-induced phase shifts in the feedback loop are incorporated into these parameters. Accordingly, due to the practically unavoidable losses, $r_B$ is bounded by $0\leq r_B<1$ and characterizes the feedback efficiency of the output field. Furthermore, the vacuum input noise introduced by the imperfections in the feedback loop has been included into a single equivalent loss channel associated with the right photonic reservoir, which is described by $\hat{a}_{2}^{\rm in,\prime}$. Therefore, by replacing the bare input noise operator $\hat{a}_{2}^{\rm in}$ in Eq.\,(\ref{eq:linQLEs}) with $\hat{a}_{2,fb}^{\rm in}$, the feedback-modified QLE for the cavity mode is obtained as
\begin{align}
\delta\dot{\hat{a}}=&-(i\tilde\Delta_{a}+\tilde{\kappa}_a)\delta\hat{a}-iG_{a}\delta\hat{x}-i\lambda\delta\hat{f}+\sqrt{2\tilde{\kappa}_a}\hat{A}^{\rm in},
\end{align}
where $\tilde{\Delta}_a = \bar\Delta_{a} - 2\sqrt{\kappa_1\kappa_2}r_{B}\sin\theta$ and $\tilde{\kappa}_{a} = \kappa_a - 2\sqrt{\kappa_1\kappa_2}r_{B}\cos\theta$ are the feedback-modified cavity detuning and decay rate, respectively. Moreover, $\hat A^{\rm in}$ denote the feedback-modified input noise operator and is given by
\begin{align}
\hat A^{\rm in} = \dfrac{1}{\sqrt{\tilde{\kappa}_a}}\left[\left(\sqrt{\kappa_{1}}-\sqrt{\kappa_{2}}r_{B}e^{i\theta}\right) \hat{a}_{1}^{\mathrm{in}}+\sqrt{\kappa_{2}}t_{B}\hat{a}_{2}^{\rm in,\prime}\right],
\end{align}
which describes the total vacuum input noises of the cavity mode and satisfies the following correlation function: $\langle\hat A^{\rm in}(t)\hat A^{\rm in,\dagger}(t')\rangle=\delta(t-t')$.

By defining the quadrature operators for the cavity and Fano modes as
\begin{align}
\delta {\hat{X}}_a&=\frac{1}{\sqrt{2}}(\delta\hat{a}+\delta\hat{a}^\dagger),
\quad \delta {\hat{Y}}_a=\frac{i}{\sqrt{2}}(\delta\hat{a}^\dagger-\delta\hat{a}),\notag\\
\delta {\hat{X}}_f&=\frac{1}{\sqrt{2}}(\delta\hat{f}+\delta\hat{f}^\dagger),
\quad \delta {\hat{Y}}_f=\frac{i}{\sqrt{2}}(\delta\hat{f}^\dagger-\delta\hat{f}),
\label{eq:quadrature1}
\end{align}
and the corresponding Hermitian input noise operators as
\begin{align}
\hat{X}_{a}^{\rm in}=&\frac{1}{\sqrt{2}}(\hat{A}^{\rm in,\dagger}+\hat{A}^{\rm in}),
\quad \hat{Y}_{a}^{\rm in}=\frac{i}{\sqrt{2}}(\hat{A}^{\rm in,\dagger}-\hat{A}^{\rm in}), \notag \\
\hat{X}_{f}^{\mathrm{in}}=&\frac{1}{\sqrt{2}}(\hat{a}_{1}^{\mathrm{in},\dagger}+\hat{a}_{1}^{\mathrm{in}}),
\quad \hat{Y}_{f}^{\mathrm{in}}=\frac{i}{\sqrt{2}}(\hat{a}_{1}^{\mathrm{in},\dagger}-\hat{a}_{1}^{\mathrm{in}}),
\label{eq:quadrature2}
\end{align}
the feedback-modified linearized QLEs for the quadrature fluctuation vector, $\hat{u}(t) = [\delta \hat{x}, \delta \hat{p}, \delta \hat{X}_a, \delta \hat{Y}_a, \delta \hat{X}_f, \delta \hat{Y}_f]^{T}$, can be written in a compact form as
\begin{align}
\dot{u}(t)=A\hat{u}(t)+\hat{n}(t),
\label{eq:LQLEs_compact}
\end{align}
where $\hat{n}(t)=[0,\hat{\xi},\sqrt{2\tilde\kappa}\hat{X}_{a}^{\rm in},\sqrt{2\tilde\kappa}\hat{Y}_{a}^{\rm in},\sqrt{2\kappa_f}{\hat{X}}_{f,1}^{\mathrm{in}},\sqrt{2\kappa_f}{\hat{Y}}_{f,1}^{\mathrm{in}}]^{T}$ is the vector of input noises, and the drift matrix A is given by
\begin{align}
A=\left(\begin{matrix}
0 & \Omega_m & 0 & 0 & 0 & 0 \\
-\Omega_m & -\gamma_m & -\sqrt{2}G_a & 0 &-\sqrt{2} G_f & 0\\
0 & 0 & -\tilde{\kappa}_{a} & \tilde{\Delta}_{a} &-\kappa_{1f} & J \\
-\sqrt{2}G_a& 0 &  -\tilde{\Delta}_{a}& -\tilde{\kappa}_{a} & -J & -\kappa_{1f} \\
0 & 0 & -\kappa_{1f} & J & -\kappa_{f} & \bar{\Delta}_{f} \\
-\sqrt{2}G_{f} & 0 & -J & -\kappa_{1f} & -\bar{\Delta}_{f} & -\kappa_{f}
\end{matrix}\right).
\label{eq:Amat}
\end{align}
The system dynamics governed by linearized QLEs (\ref{eq:LQLEs_compact}) is stable and reaches its steady state if and only if all of the eigenvalues of the drift matrix $A$ have negative real parts. Within the stable parameter regime, owing to the linearized dynamics and the Gaussian nature of the input noises, the steady state of the system is a three-mode Gaussian state. In general, the Gaussian state is completely characterized by the first and second statistical moments of the quadrature operators. Since the first moments can be arbitrarily adjusted by local unitary operations (displacements) in phase space without affecting any statistical property related to entropy or entanglement, the steady state of the system is therefore characterized by a $6\times6$ covariance matrix (CM) $V$, with the matrix entries defined as $V_{ij}=\langle{u_{i}(t)u_{j}(t^\prime)+u_{j}(t^\prime)u_{i}(t)}\rangle/2~(i,j=1,2,...6)$. When all the stability conditions are fulfilled, the steady-state CM $V$ is determined by the Lyapunov equation
\begin{align}
AV+VA^{T}=-D,
\label{eq:lyapunov}
\end{align}
where
\begin{equation}
D = \begin{pmatrix}
0 & 0 & 0 & 0 & 0 & 0 \\
0 & \gamma_m(2\bar{n}_m+1) & 0 & 0 & 0 & 0 \\
0 & 0 & \tilde{\kappa}_{a} & 0 & \alpha & \beta \\
0 & 0 & 0 & \tilde{\kappa}_{a} & -\beta & \alpha \\
0 & 0 & \alpha & -\beta & \kappa_f & 0 \\
0 & 0 & \beta & \alpha & 0 & \kappa_f
\end{pmatrix}
\label{eq:diffusion}
\end{equation}
is the diffusion matrix defined through $D_{ij}\delta(t-t^\prime)=\langle n_i(t)n_j(t') + n_j(t')n_i(t)\rangle/2~(i,j=1,2,...6)$, with
\begin{align}
\alpha &= \sqrt{\kappa_1\kappa_f} - \sqrt{\kappa_2\kappa_f}r_{B}\cos\theta, \notag\\
\beta &= \sqrt{\kappa_2\kappa_f}r_{B}\sin\theta.
\label{eq:diffusion_params}
\end{align}
The Lyapunov equation (\ref{eq:lyapunov}) is linear in $V$ and can be solved numerically, which allows us to find $V$ for any values of the relevant parameters.
Regarding the verification of bipartite entanglement, we adopt the logarithmic negativity $E_{\mathcal{N}}$ as a quantitative entanglement measure, which is defined based on specifying the positivity of the partial transpose of the CM $V$. For the continuous-variable (CV) Gaussian state, the logarithmic negativity $E_{\mathcal{N}}^{\mu|\nu}$ is defined as~\cite{Adesso2004PRA}
\begin{align}
E_{\mathcal{N}}^{\mu|\nu}=\max[0,-\ln(2\eta_{0}^{-})],\label{eq:En}
\end{align}
where $\eta_{0}^{-}\!\equiv\!2^{-1/2}\{\Sigma(V_{\mu\nu})-[\Sigma(V_{\mu\nu})^{2}-4\det\!V_{\mu\nu}]^{1/2}\}^{1/2}$ is the minimum symplectic eigenvalue of the partial transpose of a reduced $4\times4$ CM $V_{\mu\nu}$, with $\Sigma(V_{\mu\nu})\!\equiv\!\det\mathcal{A}_{\mu}+\det\mathcal{B}_{\nu}-2\det\mathcal{C}_{\mu\nu}$. The reduced CM $V_{\mu\nu}$ contains the entries of $V$ associated with the two selected modes $\mu$ and $\nu$, which can be directly obtained by removing the rows and columns of the irrelevant modes in $V$. The $2\times2$ block form of CM $V_{\mu\nu}$ is given by
\begin{align}
V_{\mu\nu}=\left(\begin{matrix}
\mathcal{A}_{\mu}&\mathcal{C}_{\mu\nu}\\
\mathcal{C}_{\mu\nu}^{\textit{T}}&\mathcal{B}_{\nu}
\end{matrix}\right).
\label{reducedCM}
\end{align}
Note that as an entanglement monotone~\cite{Plenio2005PRL}, the logarithmic negativity $E_\mathcal{N}$ quantifies how much the positivity of the partial transpose condition for separability is violated for the considered CV Gaussian states, which provides a quantitative entanglement measure based on Simon's necessary and sufficient entanglement nonpositive partial transpose criterion (or the related Peres-Horodecki criterion) for certifying bipartite Gaussian entanglement~\cite{Simon2000PRL}. Consequently, a nonzero value of $E_{\mathcal{N}}$, corresponding to $\eta_{0}^{-}<1/2$, confirms the presence of bipartite entanglement between the two selected modes $\mu$ and $\nu$.

\section{Enhancing entanglement via the synergistic control of Fano-cavity coupling, coherent feedback, and frequency mismatch}\label{secIII}
\begin{figure*}[htbp]
	\centering
	\includegraphics[width=0.98\textwidth]{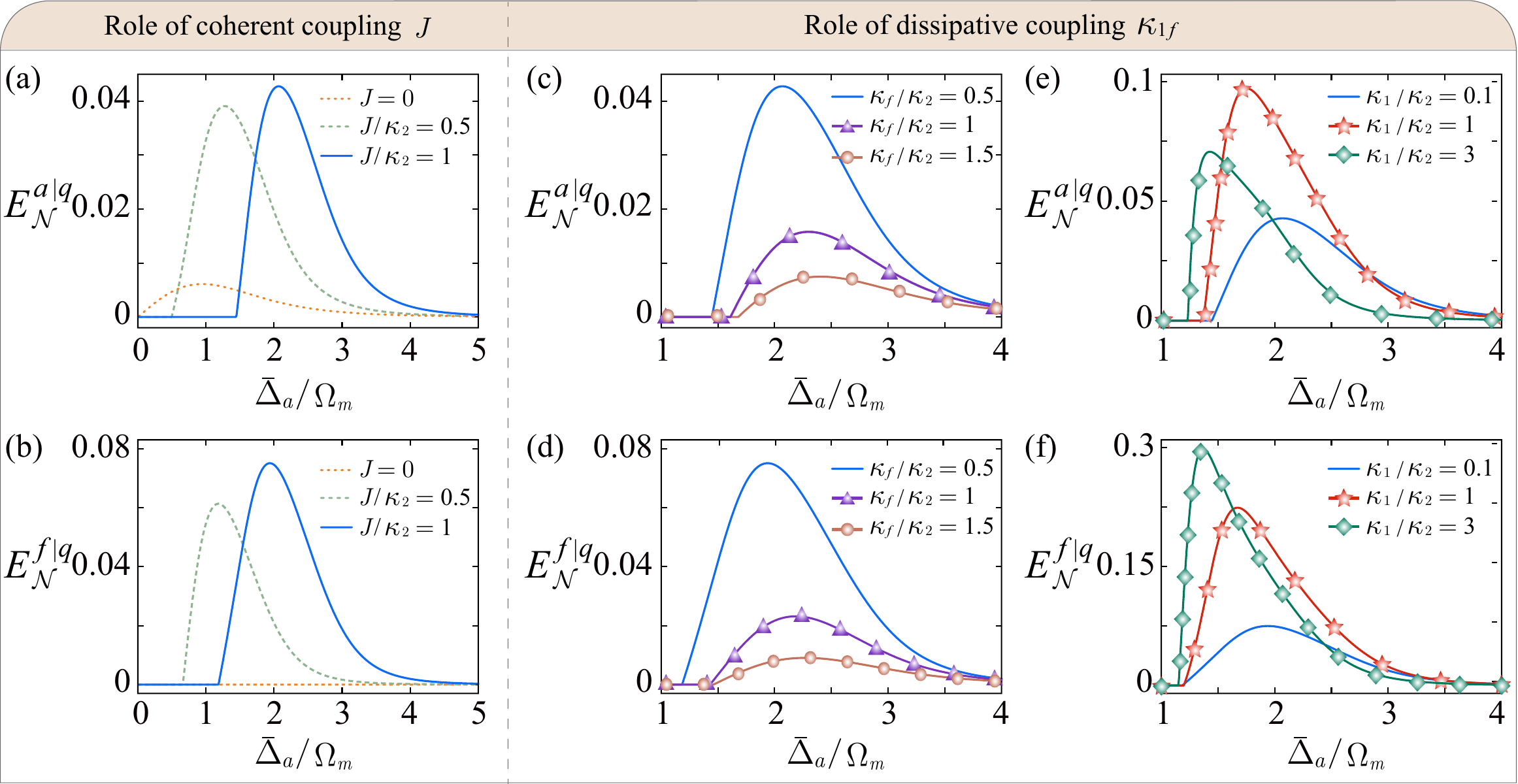}
\caption{Coherent and dissipative Fano-cavity coupling effects on optomechanical entanglement. (a,b) The logarithmic negativities $E_{\mathcal{N}}^{a|q}$ and $E_{\mathcal{N}}^{f|q}$ versus the effective cavity detuning $\bar{\Delta}_{a}/\Omega_{m}$ for different coupling strengths $J/\kappa_{2}=0,\,0.5,\,1$, with $\kappa_{1}/\kappa_{2}=0.1$ and $\kappa_{f}/\kappa_{2}=0.5$, demonstrating the enhancement of both cavity-mechanical and Fano-mechanical entanglement by coherent Fano-cavity coupling. Panels (c)-(f) show the influence of dissipative Fano-cavity coupling $\kappa_{1f}=\sqrt{\kappa_{1}\kappa_{f}}$ on different optomechanical entanglement. (c,d) $E_{\mathcal{N}}^{a|q}$ and $E_{\mathcal{N}}^{f|q}$ versus the effective cavity detuning $\bar{\Delta}_{a}/\Omega_{m}$ for different values of the Fano-mode decay rate $\kappa_{f}/\kappa_{2}=0.5,\,1,\,1.5$, with $\kappa_{1}/\kappa_{2}=0.1$ and $J/\kappa_{2}=1$. (e,f) $E_{\mathcal{N}}^{a|q}$ and $E_{\mathcal{N}}^{f|q}$ versus $\bar{\Delta}_{a}/\Omega_{m}$ for different values of the Fano-mirror decay rate $\kappa_{1}/\kappa_{2}=0.1,\,1,\,3$, with $\kappa_{f}/\kappa_{2}=0.5$ and $J/\kappa_{2}=1$. Although both $\kappa_1$ and $\kappa_f$ increase $\kappa_{1f}$, $\kappa_f$ monotonically suppresses entanglement, whereas $\kappa_1$ yields a nonmonotonic response. The other parameters used here are provided in the text.}\label{Fig2}
\end{figure*}

In Fig.\,\ref{Fig2}, in order to characterize the optomechanical entanglement in a Fano-mirror COM system, we plot the logarithmic negativity $E_{\mathcal{N}}^{a|q}$ (cavtiy-mechanical entanglement) and $E_{\mathcal{N}}^{f|q}$ (Fano-mechanical entanglement) as functions of the effective cavity detuning $\bar{\Delta}_{a}/\Omega_{m}$, with both of the coherent and dissipative Fano-cavity couplings being considered. In the numerical calculation, to ensure the system stability and experimental feasibility, the following parameters are employed~\cite{Thompson2008Nature}: $\kappa_{2}/2\pi=2\,\textrm{MHz}$, $\Omega_{m}/2\pi=1.3\,\textrm{MHz}$, $\gamma_{m}/2\pi=0.12\,\textrm{Hz}$, $g_{a,0}/\Omega_{m}=6.5\times10^{-5}$, $g_{f,0}/\Omega_{m}=-1.6\times10^{-4}$, $\lambda_{d}=810\,\textrm{nm}$, $P_{d}=0.18\,\textrm{mW}$, $T=400\,\textrm{mK}$, and optical frequency mismatch $\delta\Delta\equiv\Delta_a-\Delta_f=\omega_a-\omega_f=0$. As shown in Figs.~\ref{Fig2}(a) and \ref{Fig2}(b), we start by investigating the dependence of $E_{\mathcal{N}}^{a|q}$ and $E_{\mathcal{N}}^{f|q}$ on coherent Fano-cavity coupling strength $J$, with the dissipative Fano-cavity coupling $\kappa_{1f}=\sqrt{\kappa_{1}\kappa_{f}}$ fixed by choosing $\kappa_{1}/\kappa_{2}=0.1$ and $\kappa_{f}/\kappa_{2}=0.5$. In the absence of the Fano mode, corresponding to a conventional COM system with $J=0$ and $g_{f,0}=0$, $E_{\mathcal{N}}^{a|q}$ is nonvanishing within a finite parameter region around the COM resonance $\bar{\Delta}_{a}\simeq\Omega_{m}$ with a maximum value of about $0.006$, whereas $E_{\mathcal{N}}^{f|q}$ is identically zero throughout the considered detuning range. When the Fano mode is introduced into the COM system, i.e., $J\neq0$ and $g_{f,0}\neq0$, both $E_{\mathcal{N}}^{a|q}$ and $E_{\mathcal{N}}^{f|q}$ become nonzero and increase monotonically with the coherent Fano-cavity coupling strength $J$, accompanied by a shift of their peak position toward the blue-detuned regime, with maximum peak values of $E_{\mathcal{N}}^{a|q}\simeq0.04$ and $E_{\mathcal{N}}^{f|q}\simeq0.08$, respectively. These results indicate that the introduction of the Fano mode not only gives rise to a new and stronger type of Fano-mechanical entanglement, but also provides a coherent Fano-cavity coupling that can further enhance both types of COM entanglement. In Figs.~\ref{Fig2}(c)-\ref{Fig2}(f), we further examine the influence of the dissipative Fano-cavity coupling $\kappa_{1f}$ on COM entanglement by varying the two optical decay rates $\kappa_1$ and $\kappa_f$. It is seen that while increasing either $\kappa_1$ or $\kappa_f$ serves to enhance the dissipative coupling $\kappa_{1f}$, these two mechanisms exhibit markedly different impacts on both of cavity-mechanical and Fano-mechanical entanglements. Specifically, as shown in Figs.\,\ref{Fig2}(c) and \ref{Fig2}(d), we plot $E_{\mathcal{N}}^{a|q}$ and $E_{\mathcal{N}}^{f|q}$ as functions of $\bar{\Delta}_{a}/\Omega_{m}$ for different values of $\kappa_{f}$, with $\kappa_{1}/\kappa_{2}=0.1$ and $J/\kappa_{2}=1$. It is seen that as the Fano-mode decay rate $\kappa_{f}$ increases, the maximum values of $E_{\mathcal{N}}^{a|q}$ and $E_{\mathcal{N}}^{f|q}$ are directly suppressed. Figures\,\ref{Fig2}(e) and \ref{Fig2}(f) further show $E_{\mathcal{N}}^{a|q}$ and $E_{\mathcal{N}}^{f|q}$ as functions of $\bar{\Delta}_{a}/\Omega_{m}$ for different values of $\kappa_{1}$, with $\kappa_{f}/\kappa_{2}=0.5$ and $J/\kappa_{2}=1$. In this case, it is found that with the increase of the Fano-mirror decay rate $\kappa_{1}$, $E_{\mathcal{N}}^{a|q}$ first increases and then decreases, attaining a maximum value of about $0.1$ for $\kappa_{1}/\kappa_{2}=1$. This maximum value is enhanced by approximately $16$ times compared to the scheme without the Fano mode ($J=0$ and $g_{f,0}=0$), and by nearly $2.5$ times compared with the weak decay rate case ($\kappa_{1}/\kappa_{2}=0.1$). In contrast, $E_{\mathcal{N}}^{f|q}$ increases monotonically with the Fano-mirror decay rate $\kappa_{1}$, achieving a maximum value of about $0.3$, which is enhanced by a factor of $3.7$ compared with the weak decay rate case ($\kappa_{1}/\kappa_{2}=0.1$).
\begin{figure*}[t]
	\centering
	\includegraphics[width=0.98\textwidth]{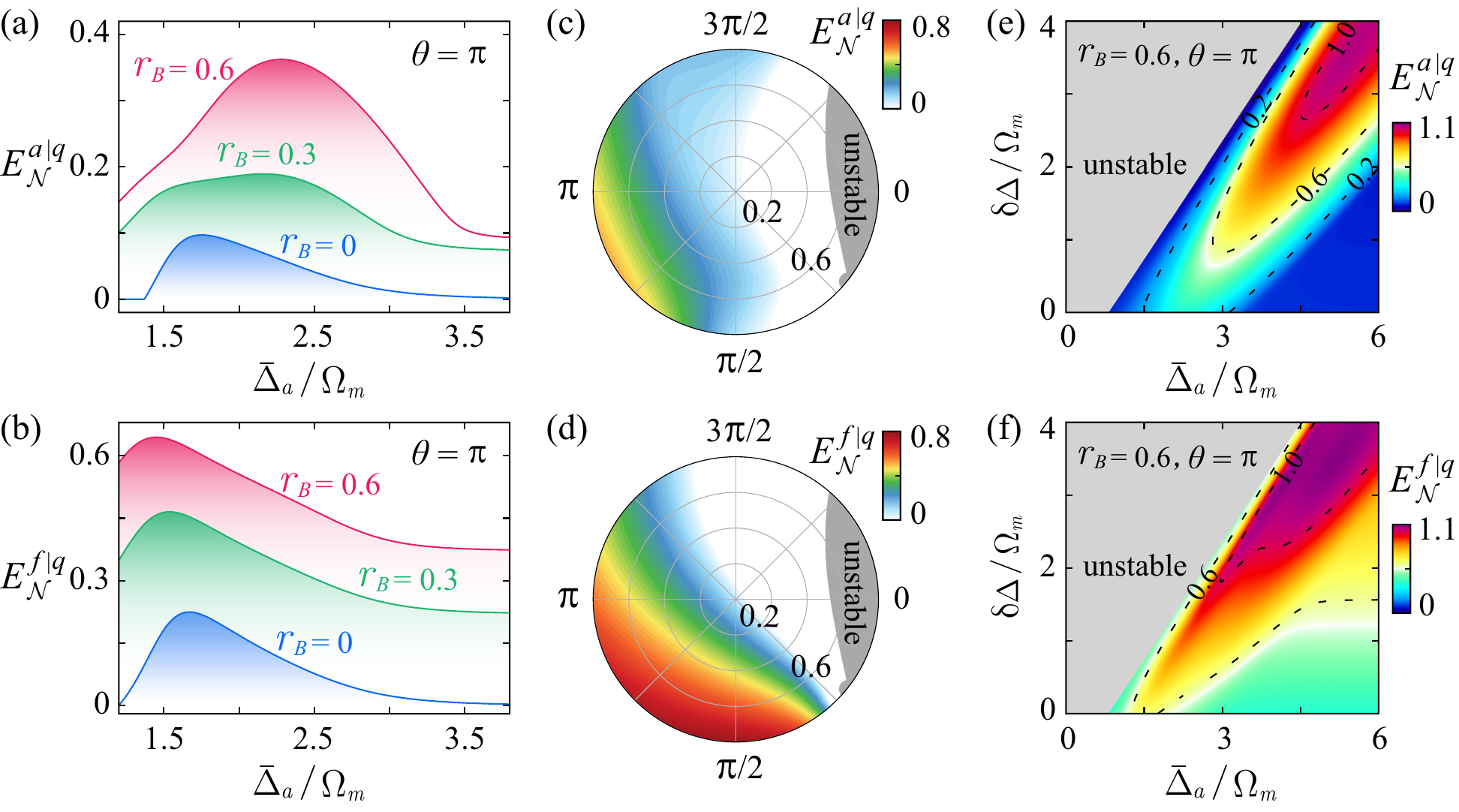}
\caption{Enhancing optomechanical entanglement in Fano-mirror COM system via the synergistic control of coherent feedback and optical frequency mismatch. (a,b) The logarithmic negativities $E_{\mathcal{N}}^{a|q}$ and $E_{\mathcal{N}}^{f|q}$ are plotted as a function of the effective cavity detuning $\bar{\Delta}_{a}/\Omega_{m}$ for different values of the feedback strength $r_{B}=0,\,0.3,\,0.6$, with $\theta=\pi$ and $\delta\Delta=0$. (c,d) Polar plots of the logarithmic negativities $E_{\mathcal{N}}^{a|q}$ and $E_{\mathcal{N}}^{f|q}$ as functions of the feedback strength $r_{B}$ (radial coordinate) and the feedback phase $\theta$ (angular coordinate), with $\bar{\Delta}_{a}/\Omega_{m}=2.5$ and $\delta\Delta=0$. (e,f) Density plots of the logarithmic negativities $E_{\mathcal{N}}^{a|q}$ and $E_{\mathcal{N}}^{f|q}$ as functions of the effective cavity detuning $\bar{\Delta}_{a}/\Omega_{m}$ and the optical detuning difference $\delta\Delta$, with $r_{B}=0.6$ and $\theta=\pi$.
These results show that the synergistic control provide an efficient means for enhancing optomechanical entanglement. Here, we take $\kappa_{1}/\kappa_{2}=1$, $\kappa_{f}/\kappa_{2}=0.5$, $J/\kappa_{2}=1$, and the other parameters are the same as in Fig.\,\ref{Fig2}.}\label{Fig3}
\end{figure*}

After having explored the unique feature of Fano-cavity coupling on the generation and manipulation of COM entanglement, we now investigate how to further enhance such entanglement via the injection of coherent feedback. To this end, as shown in Fig.\,\ref{Fig3}, we plot the dependence of the logarithmic negativity $E_{\mathcal{N}}^{a|q}$ and $E_{\mathcal{N}}^{f|q}$ on the feedback strength $r_{B}$, the feedback phase $\theta$, and the optical detuning difference $\delta\Delta$. For the numerical calculation, we here fix the coherent and dissipative Fano-cavity coupling strength by setting $J/\kappa_{2}=1$, $\kappa_{1}/\kappa_{2}=1$, and $\kappa_{f}/\kappa_{2}=0.5$, while keeping the other parameters the same as those used in Fig.\,\ref{Fig2}. Figures\,\ref{Fig3}(a) and \ref{Fig3}(b) show the dependence of $E_{\mathcal{N}}^{a|q}$ and $E_{\mathcal{N}}^{f|q}$ on the feedback strength $r_{B}$ under the condition of $\theta=\pi$ and $\delta\Delta=0$. Here, $r_{B}=0$ corresponds to the case of $\kappa_{1}/\kappa_{2}=1$ (red star lines) in Figs.\,\ref{Fig2}(e) and \ref{Fig2}(f), serving as a reference case without feedback control. In the presence of feedback, i.e., $r_{B}\neq0$, both of $E_{\mathcal{N}}^{a|q}$ and $E_{\mathcal{N}}^{f|q}$ can be enhanced when the feedback strength $r_{B}$ increases, characterized by an amplified maximum value as well as a broadening of the detuning range over which COM entanglement is generated. For example, in the case of $r_{B}=0.6$, the maximum values of $E_{\mathcal{N}}^{a|q}$ and $E_{\mathcal{N}}^{f|q}$ reach approximately $0.36$ and $0.64$, respectively, corresponding to an enhancement of approximately $3$ times compared with the case without feedback. In Figs.\,\ref{Fig3}(c) and \ref{Fig3}(d), we further present the dependence of $E_{\mathcal{N}}^{a|q}$ and $E_{\mathcal{N}}^{f|q}$ on the feedback phase $\theta$ by polar plots, in which $r_{B}$ and $\theta$ serve as the radial and angular coordinates, respectively, with $\bar{\Delta}_{a}/\Omega_{m}=2.5$ and $\delta\Delta=0$. It is found that both $E_{\mathcal{N}}^{a|q}$ and $E_{\mathcal{N}}^{f|q}$ exhibit a periodic dependence on $\theta$ with a period of $2\pi$. The two types of COM entanglement emerge roughly within the phase range from $\theta=\pi/4$ to $\theta=3\pi/2$, within which the magnitude of entanglement increases monotonically with $r_{B}$. In contrast, for the remaining phase ranges, both of $E_{\mathcal{N}}^{a|q}$ and $E_{\mathcal{N}}^{f|q}$ remain absent even with increasing feedback strength, and a sufficiently large value of $r_{B}$ could instead cause the system to enter the unstable parameter regime. Meanwhile, it is seen that within the phase range where entanglement is present, both $E_{\mathcal{N}}^{a|q}$ and $E_{\mathcal{N}}^{f|q}$ exhibit an optimal phase $\theta$ at which their respective maximum values are attained. For instance, in the case of $r_{B}=0.8$, $E_{\mathcal{N}}^{a|q}$ reaches a maximum value of about $0.5$ at $\theta=4\pi/5$, whereas $E_{\mathcal{N}}^{f|q}$ attains a maximum value of about $0.78$ at $\theta=\pi/2$. Figures\,\ref{Fig3}(e) and \ref{Fig3}(f) show the dependence of $E_{\mathcal{N}}^{a|q}$ and $E_{\mathcal{N}}^{f|q}$ on $\bar{\Delta}_{a}/\Omega_{m}$ and $\delta\Delta$, respectively, with $r_{B}=0.6$ and $\theta=\pi$. Interestingly, it is found that a finite detuning difference between the cavity and Fano modes ($\delta\Delta>0$) can further enhance both types of COM entanglement in the presence of coherent feedback. In the case of $r_{B}=0.6$ and $\theta=\pi$, the maximum values of $E_{\mathcal{N}}^{a|q}$ and $E_{\mathcal{N}}^{f|q}$ reach about $1.05$ and $1.1$, respectively, corresponding to approximately $3$ times and $1.7$ times their respective values in the cavity-Fano resonant case ($\delta\Delta=0$). However, it is also seen that this enhancement is accompanied by an expansion of the unstable parameter region with increasing $\delta\Delta$, which narrows the detuning range over which the entanglement can be sustained.

The results shown in Figs.\,\ref{Fig2} and \ref{Fig3} indicate that the introduction of the Fano mirror and the coherent feedback loop not only enriches the types of bipartite entanglement that can be generated, but also significantly enhances the magnitude of the original cavity-mechanical entanglement. Physically, these behaviors can be understood as follows. In our proposed scheme, the original cavity-mechanical entanglement arises from the two-mode squeezing interaction mediated by the direct optomechanical coupling between the cavity and mechanical modes. When a Fano mirror is introduced, the auxiliary Fano mode couples simultaneously to the cavity and mechanical modes, forming a closed tripartite coupling loop. This closed-loop interaction structure enables an additional Fano-mediated indirect coupling between cavity and mechanical modes, which allows the associated cavity-mechanical entanglement to be enhanced through the quantum interference between the direct and indirect pathways. Here, it is worth noting that, although increasing both coherent and dissipative Fano-cavity couplings can strengthen the quantum interference in the closed interaction loop and thereby enhance cavity-mechanical entanglement, an excessively large dissipative coupling also tends to introduce additional loss that speeds up decoherence and suppresses the entanglement. This competing behavior gives rise to the nonmonotonic dependence of $E_{\mathcal{N}}^{a|q}$ shown in Fig.~\ref{Fig2}(e). Interestingly, the coherent feedback loop provides an effective approach to overcome this problem through simultaneously renormalizing the effective cavity detuning and damping and reshaping their input noise properties. Specifically, since the cavity and Fano modes couple to the same photonic reservoir at the left cavity boundary, feedback control can thus enhance the input-noise correlations between the two optical modes [see Eqs.~(\ref{eq:diffusion}) and (\ref{eq:diffusion_params})], thereby strengthening the interference between the direct and Fano-mediated indirect interaction pathways. Meanwhile, an appropriate frequency mismatch between the two optical modes can also compensate for the Fano-cavity detuning shift induced by the feedback loop, leading to further optimization of the interference condition. In summary, through properly tuning the Fano-cavity coupling and the synergistic control of coherent feedback and optical frequency mismatch, the original cavity-mechanical entanglement can be remarkably enhanced, as well as additional Fano-mechanical entanglement is generated.

\section{Predicting quantum entanglement based on machine learning}\label{secIV}
Up to now, the bipartite entanglement measures $E_{\mathcal{N}}^{a|q}$ and $E_{\mathcal{N}}^{f|q}$ discussed above are obtained by numerically solving the Lyapunov equation (\ref{eq:lyapunov}) for the steady-state CM. In experiments, however, such entanglement measures are not directly measurable, because certifying bipartite entanglement requires the reconstruction of the full three-mode CM, which in turn demands the simultaneous detection of all the optical and the mechanical quadratures. Since the mechanical quadratures are difficult to access directly and the reconstruction is sensitive to detection noise and parameter fluctuations, it is desirable to develop a more practical route for estimating such COM entanglement. To this end, we adopt a supervised machine learning approach, in which a feedforward neural network is trained to predict the entanglement measures directly from the readily measurable optical signals and the feedback parameters, which is inspired by the machine-learning scheme proposed for predicting strong phonon blockade~\cite{Zeng2021APL}. The feasibility of such an approach is supported by the universal approximation theorem, which states that a feedforward network with a single hidden layer and a finite number of neurons can approximate continuous functions on compact subsets under mild assumptions on the activation function~\cite{Hornik1991NN}, and by the wide applications of machine learning techniques in dealing with complex quantum problems~\cite{Carleo2019RMP}.
\begin{figure*}[t]
	\centering
	\includegraphics[width=0.98\textwidth]{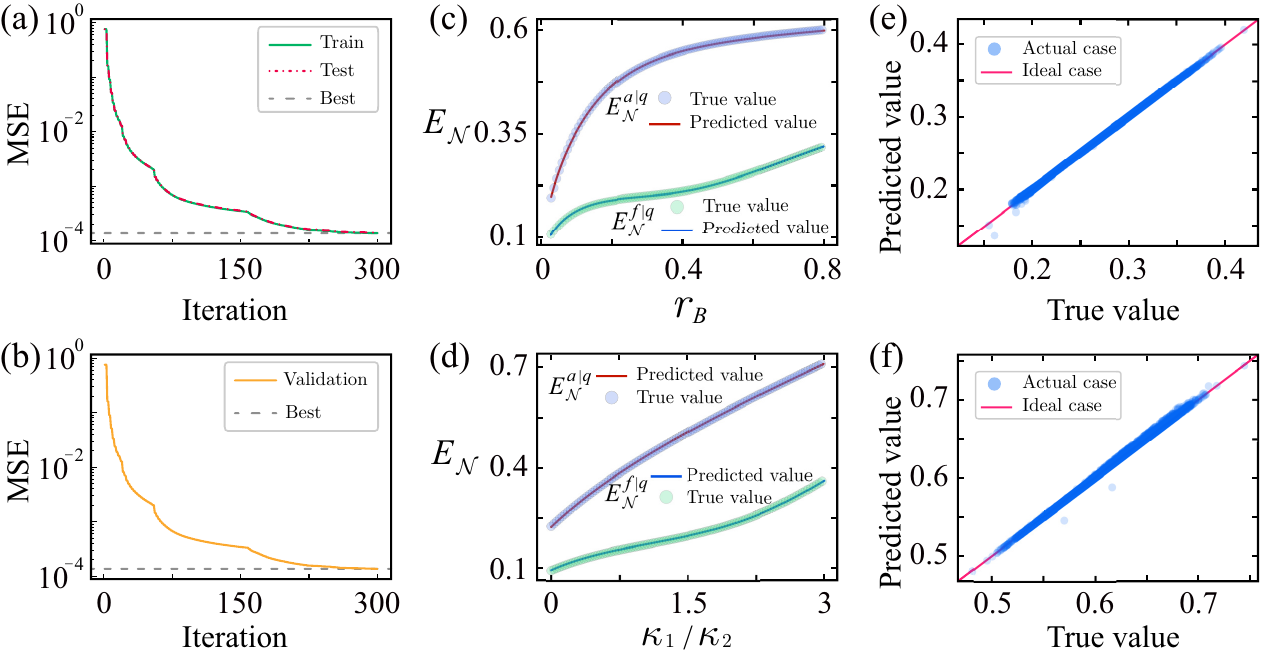}
\caption{Training process of the feedforward neural network and validation of the trained model through numerical experiments. (a) The mean squared error MSE for the training set and test set versus the iterations. (b) The mean squared error MSE for the validation set versus the iterations. The whole sample set includes $1\times 10^{6}$ examples where $70\%$ of the whole sample set is the training set, $15\%$ is used as the validation set, and the remaining  $15\%$  is considered as the test samples. (c) The real and predicted values of $ E_{\mathcal{N}}^{a|q}$ and $E_{\mathcal{N}}^{f|q}$ versus $r_B$ with $\kappa_{1}/\kappa_{2}=1$. (d) The real and predicted values of $ E_{\mathcal{N}}^{a|q}$ and $E_{\mathcal{N}}^{f|q}$ versus $\kappa_{1}/\kappa_{2}$ with $r_{B}=0.4$. (e,f) The predicted values of $ E_{\mathcal{N}}^{a|q}$ and $ E_{\mathcal{N}}^{f|q}$ versus their real values in the case of $(r_B,\kappa_{1}/\kappa_{2},{\bar\Delta_a}/\Omega_m)$ being random numbers with small fluctuations $(0.1\xi_{1}, 0.1\xi_{2}, 0.35\xi_{3})$ around $(0.4, 1, 1)$. The number of random numbers is $20000$. The noises $\xi_{1}$, $\xi_{2}$ and $\xi_{3}$ obey the Gaussian distribution $N(0,0.3)$. Here, we take $J/\kappa_{2}=1$, $\bar{\Delta}_{a}/\Omega_{m}=\bar{\Delta}_{f}/\Omega_{m}=1.7$, $\theta=\pi$, and the other parameters used here are the same as in Fig.\,\ref{Fig2}.}\label{Fig4}
\end{figure*}

In the $m$-th sample of the supervised learning task, the input vector is chosen as
\begin{align}
X^{m}=&\{X_a^{(m)},Y_a^{(m)},X_f^{(m)},Y_f^{(m)},\notag\\
&\;\;(r_B\sin\theta)^{(m)},(r_B\cos\theta)^{(m)}\},\label{eq:input}
\end{align}
where the first four components are the steady-state mean values of the cavity and Fano quadratures, which can be measured by balanced homodyne detection shown in Fig.\ref{Fig1}(a), and the last two components encode the feedback amplitude $r_B$ and phase $\theta$ through their trigonometric projections, so that the $2\pi$-periodic phase is represented in a continuous manner. The corresponding output is the two-component vector $Y^{m}=\{E_{\mathcal{N}}^{a|q},E_{\mathcal{N}}^{f|q}\}_{m}$, whose ideal values are computed from the steady-state CM by solving the Lyapunov equation (\ref{eq:lyapunov}) together with Eq.~(\ref{eq:En}). The whole sample set is defined as
\begin{align}
S=\left\{(X^{1},Y^{1}),(X^{2},Y^{2}),\ldots,(X^{N},Y^{N})\right\},\label{eq:sampleset}
\end{align}
where $N$ is the number of samples. The sample set is partitioned into a training set $S_{\mathrm{train}}$, a validation set $S_{\mathrm{val}}$, and a test set $S_{\mathrm{test}}$, i.e., $S=S_{\mathrm{train}}\cup S_{\mathrm{val}}\cup S_{\mathrm{test}}$, which are utilized to estimate the model, to control the complexity of the network, and to evaluate the performance of the finally selected optimal model, respectively. We adopt a feedforward neural network whose input layer has six nodes corresponding to the six components of $X^{m}$ and whose output layer has two nodes corresponding to the two entanglement measures in $Y^{m}$. In our simulation, the network contains two hidden layers, each with $32$ neurons, the hyperbolic tangent function $\tanh(x)$ is adopted as the activation function, and the weight matrix and bias vector are optimized by the Levenberg-Marquardt algorithm, following the configuration in Ref.~\cite{Zeng2021APL}. The training task is to optimize the weight matrix and bias vector such that the error between the ideal output $Y^{k}$ and the estimated output $Y^{\prime k}$ of the network is minimized. The performance of the neural network is reflected by the mean squared error (MSE), which is calculated via
\begin{align}
\mathrm{MSE}=\frac{1}{N}\sum_{k=1}^{N}\left(Y^{k}-Y^{\prime k}\right)^{T}\left(Y^{k}-Y^{\prime k}\right),\label{eq:MSE}
\end{align}
where $N$ denotes the number of samples in the corresponding (training, validation, or test) set. The iteration number of the training is determined by monitoring the MSE of the test set, and the network with the smallest test-set MSE is retained as the final model for the subsequent predictions.

To train and validate the network, we generate a total of $1\times10^{6}$ samples by randomly sampling the system parameters within the stable parameter regime and solving the Lyapunov equation (\ref{eq:lyapunov}) for the corresponding CM, from which the ideal values of $E_{\mathcal{N}}^{a|q}$ and $E_{\mathcal{N}}^{f|q}$ are obtained. The whole sample set is divided into a training set ($70\%$), a validation set ($15\%$), and a test set ($15\%$). The training process is illustrated in Figs.\,\ref{Fig4}(a) and \ref{Fig4}(b), where the MSEs of the training and test sets and that of the validation set are plotted versus the iteration number. It is seen that all the MSEs decrease rapidly within the first iterations and then saturate at a small value of about $10^{-4}$, with the optimal iterations labeled by ``Best'' (gray dashed lines). The monotonic convergence of the MSEs in Fig.\,\ref{Fig4}(a) shows that the training target is achieved, and the small validation-set MSE in Fig.\,\ref{Fig4}(b) indicates that the chosen network structure and the optimized parameters are appropriate.

In Figs.\,\ref{Fig4}(c) and \ref{Fig4}(d), to demonstrate the predictive performance of the trained neural network, we compare the predicted values with the true values of $E_{\mathcal{N}}^{a|q}$ and $E_{\mathcal{N}}^{f|q}$ as functions of $\kappa_{1}/\kappa_{2}$ with $r_{B}=0.4$ and as functions of $r_{B}$ with $\kappa_{1}/\kappa_{2}=1$, respectively. Here, we take $J/\kappa_{2}=1$, $\bar{\Delta}_{a}/\Omega_{m}=\bar{\Delta}_{f}/\Omega_{m}=1.7$, $\theta=\pi$, and the other parameters are the same as in Fig.\,\ref{Fig2}. For each swept parameter, the corresponding steady-state optical signals are fed into the network, and the network outputs are compared with the ideal entanglement values obtained from the Lyapunov equation. It is found that the predicted curves are almost coincident with the true curves over the entire considered parameter range for both $E_{\mathcal{N}}^{a|q}$ and $E_{\mathcal{N}}^{f|q}$, without appreciable deviation. This result indicates that the trained neural network has learned the nonlinear mapping between the measured optical signals and the two bipartite entanglement measures, enabling reliable predictions across the parameter region covered by the sampled data.

Moreover, as shown in Figs.\,\ref{Fig4}(e) and \ref{Fig4}(f), one of the advantages of the neural network is its robustness against the noise induced by the slight fluctuations of the system parameters. As a specific example, the parameters $(r_{B},\kappa_{1}/\kappa_{2},\bar{\Delta}_{a}/\Omega_{m})$ are taken as random numbers with small fluctuations $(0.1\xi_{1},0.1\xi_{2},0.35\xi_{3})$ around $(0.4,1,1)$, where $\xi_{1}$, $\xi_{2}$, and $\xi_{3}$ obey the Gaussian distribution $N(0,0.3)$, and the number of random samples is $2\times10^{4}$. We plot the predicted values of $E_{\mathcal{N}}^{a|q}$ and $E_{\mathcal{N}}^{f|q}$ versus their true values, where the ideal case corresponds to the solid line. It is seen that the predicted points compactly disperse around the ideal case for both entanglement measures, showing that the trained neural network is robust against the random disturbance of the system parameters. Therefore, even when the system parameters are not precisely known, i.e., the system effectively works as a black box, one can still reliably estimate the bipartite entanglement from the measured optical signals alone.

These results discussed in Fig.\,\ref{Fig4} indicate that the machine-learning scheme provides an efficient and robust approach for predicting bipartite COM entanglement, which relaxes the requirement of the full quantum state tomography and enhances the feasibility of the entanglement characterization in current experimental platforms.

\section{Conclusion}\label{secV}

In summary, we have investigated how to generate and enhance bipartite optomechanical entanglement in a Fano-mirror COM system through the synergistic control of Fano-cavity coupling, coherent feedback, and optical frequency mismatch, while also demonstrating the machine-learning-based prediction of the generated entanglement from measurable optical signals. Specifically, we show that when introducing a Fano mirror with both coherent and dissipative Fano-cavity couplings to COM system, it forms a closed tripartite interaction loop among the cavity, Fano, and mechanical modes. Therefore, the quantum interference between the direct and indirect cavity-mechanical coupling pathways emerges, which not only substantially enhances the original cavity-mechanical entanglement, but also gives rise to a robust Fano-mechanical entanglement. Based on this unique feature, it is also found that the coherent and dissipative Fano-cavity couplings play qualitatively distinct roles in manipulating both types of COM entanglement, and that the coherent feedback loop combined with an appropriate optical frequency mismatch can further enhance them remarkably and broaden the parameter region over which they can be sustained. More interestingly, we demonstrate that both types of COM entanglement can be reliably predicted directly from the readily measurable optical signals using a trained supervised feedforward neural network, which relaxes the requirement of full quantum-state tomography and remains robust against parameter fluctuations. Although we have considered here a specific Fano-mirror COM setup, we note that the proposed synergistic scheme remains applicable to a broad range of optomechanical platforms featuring closed-loop interactions~\cite{Huang2022PRA2,Lai2022PRR,Lai2022PRL} and coherent feedback~\cite{Li2017PRA}. These findings, establishing an all-optical and experimentally feasible route to generate, enhance, and characterize macroscopic COM entanglement, could make COM system practical for numerous entanglement-enabled quantum technologies~\cite{Zhang2019NSR,Degen2017RMP,Kimble2008Nature,Dowling2003PTRSA,Aspelmeyer2014RMP,Xiong2015SCPMA,Barzanjeh2022NP}.

\section{Acknowledgment}
Y.-F.J. was supported by the the National Natural Science Foundation of China (NSFC, Grant No.~12405029) and the Natural Science Foundation of Henan Province (Grant No.~252300421221). H.J. was supported by the NSFC (Grant No.~12421005), National Key $\&$ Program of China (Grant No.~2024YFE0102400), and the Hunan Provincial Major Sci-Tech Program (Grant No.~2023ZJ1010). L.-M.K. was supported by the NSFC (Grants No. 12675017, 12247105, and 12421005), Quantum Science and Technology-National Science and Technology Major Project (Grant No. 2024ZD0301000), and the Henan Science and Technology Major Project (Grant No. 241100210400). Q.-S.T. was supported by NSFC (Grant No. 12275077).

%

\end{document}